\documentstyle[eqsecnum,prd,aps]{revtex}
\draft
\begin{document}

\def\beq{\begin{equation}}
\def\eeq{\end{equation}}
\def\beqa{\begin{eqnarray}}
\def\eeqa{\end{eqnarray}}
\def\ben{\begin{enumerate}}
\def\een{\end{enumerate}}
\def\bed{\begin{description}}
\def\eed{\end{description}}
\def\begineqn{\begin{eqnarray}}
\def\endeqn{\end{eqnarray}}
\def\p{\partial}
\def\a{\alpha}
\def\b{\beta}
\def\g{\gamma}
\def\k{\kappa}
\def\d{\delta}
\def\arrow{\rightarrow}
\def\z{\zeta}
\def\C{{\cal C}}
\def\O{\Omega}
\newcommand{\simg}
   {\mathrel{\raise.3ex\hbox{$>$\kern-.75em\lower1ex\hbox{$\sim$}}}}
\newcommand{\siml}
   {\mathrel{\raise.3ex\hbox{$<$\kern-.75em\lower1ex\hbox{$\sim$}}}}


\thispagestyle{empty}
{\baselineskip0pt
\leftline{\large\baselineskip16pt\sl\vbox to0pt{\hbox{\it Department
of Physics}
               \hbox{\it Kyoto University}\vss}}
\rightline{\large\baselineskip16pt\rm\vbox to20pt{
          \hbox{KUNS 1386}
          \hbox{KUCP 0093}
          \hbox{March 1996}
\vss}}%
}
\vskip1cm
\begin{center}{\large \bf
Critical Behavior in the Brans-Dicke Theory of Gravitation
}
\end{center}

\vskip0.5cm

\begin{center}
 {\large
Takeshi Chiba{\footnote{
e-mail: chiba@tap.scphys.kyoto-u.ac.jp}
\footnote{Address after April: 
Yukawa Institute for Theoretical Physics, Kyoto University, 
Kyoto 606-01, Japan}
}
} \\
{\em Department of Physics,~Kyoto University,} \\
{\em Kyoto 606-01,~Japan}

{\large
Jiro Soda{\footnote{e-mail: jiro@phys.h.kyoto-u.ac.jp}}
}\\
{\em Department of Fundamental Sciences,~ FIHS,}\\
{\em Kyoto University,~Kyoto 606,~Japan}
\end{center}

\begin{abstract}
The collapse of a massless scalar field in the
Brans-Dicke theory of gravitation is studied
in the analysis of both analytical solution and numerical one.
By conformally transforming the Roberts's solution into
the Brans-Dicke frame, we find for $\omega > -3/2$ that
a continuous self-similarity continues and that
the critical exponent does depend on $\omega$.
By conformally transforming the Choptuik's solution into
the Brans-Dicke frame, we find for $\omega > -3/2$ that
at the critical solution shows discrete self-similarity,
however, the critical exponent depends strongly on 
$\omega$ while the echoing parameter weakly on it.

\end{abstract}


\section{INTRODUCTION}

Critical phenomena in black hole formation found by
Choptuik\cite{chop} has renewed interest
in the classical general relativistic black hole formation.
Among the most interesting things in Choptuik's results are

\noindent (1){\it Power law}: Black hole mass exhibits a power law
\beq
M_{BH} \propto (p-p^*)^{\beta},
\eeq
where $p$ is a parameter which characterizes the strength
of initial condition and $p^*$ is the threshold value.

\noindent (2){\it Echoing}: Configurations sufficiently close to
critical show a discrete homotheticity (or scale invariance)
\beq
\phi(\rho-\Delta, \tau-\Delta) \simeq \phi(\rho, \tau),
\eeq
where $\rho$ and $\tau$ are logarithms of proper radius $r$ and
central proper time $t$. Echoing means that the features of
critical solution are repeated on ever decreasing time
length scales.

\noindent (3){\it Universality}: The exponent $\beta \simeq 0.37$
and echoing parameter $\Delta$ are independent of any choice of
initial data $p$.

It should be noted, however,  that
the universality mentioned there refers to the independence
from the initial condition of the matter field considered.
To investigate the dependence of the model of the matter,
the collapse of gravitational wave\cite{ae} and
radiation fluid collapse\cite{ec} were
examined. For vacuum gravity, $\beta \simeq 0.37$
and $\Delta \simeq 0.6$, while for radiation fluid $\beta \simeq 0.36$
with $\Delta $ being arbitrary(i.e. continuous self-similarity).
These results excite the expectation that
the critical exponent $\beta$ may be universal
among matter fields having massless property, although
quantum effects will destroy the phenomena against expectation\cite{cs}.

In this paper, we will investigate another model dependence:
{\it dependence on the theory of gravitation}.
To see this we take the Brans-Dicke theory of gravitation\cite{bd}
for its simplicity. The theory contains a parameter $\omega$
which controls the strength of gravity, and in the limit
$\omega \arrow \infty$ the theory reduces to the Einstein theory.
We consider the Brans-Dicke theory in vacuum.
The scalar field examined here is thus the Brans-Dicke scalar field.

The collapse of a massless scalar field in the Brans-Dicke theory
is also interesting in the light of structural stability of
Choptuik's solution, which is our next problem.

Our paper is organized as follows.
In sec.2, Roberts's self-similar solution in the Brans-Dicke
theory is studied to introduce the notation and the detail of
the conformal transformation.
In sec.3, basic equations and numerical procedures are given.
We reproduce Choptuik solution.
In sec.4, numerical results for Brans-Dicke theory are given.
Sec.5 is devoted to summary.

\section{Roberts's solution in Brans-Dicke theory}
Roberts\cite{rob} found solutions of the Einstein equations
which describe  continuous self-similar spherical collapse of a
massless scalar field and he studied these solutions in the context of
counter-examples to the cosmic censorship.
Subsequntly, the solutions were studied as an analytical example
of critical behavior. We shall try to find the counterparts in
the Brans-Dicke theory of Roberts's solutions  by conformal transformation.

The action of Brans-Dicke theory without matter term is
\beq
S=\int d^4x \sqrt{-\tilde{g}}(\Phi \tilde{R}-{\omega \over \Phi}(
\tilde{\nabla }\Phi)^2 ).
\label{classbd}
\eeq
By the conformal transformation
\beqa
\tilde{g}_{ab}&=&\Phi^{-1}g_{ab},\nonumber\\
\Phi&=&\exp(\phi/\sqrt{2\omega +3}),
\label{conf}
\eeqa
the action Eq.(\ref{classbd}) is reduced to that of a
massless scalar field coupled to
the Einstein gravity where the solutions are known:
\beq
S=\int d^4x \sqrt{-g}(R-{1\over 2}(\nabla\phi)^2 ).
\label{class1}
\eeq
Therefore we simply replace the Roberts's solutions by Eq.(\ref{conf})
to find the counterparts in the Brans-Dicke theory.

The Roberts's solutions are written as
\beqa
ds^2&=&-dudv +{1\over 4}(u-(1-p)v)(u-(1+p)v)d\Omega^2,\\
\phi&=&{\pm}\Bigl(\log{u-(1+p)v\over u-(1-p)v}\Bigr),
\eeqa
where $p>0$ is an arbitrary constant.
Then those in the Brans-Dicke theory are given by
\beqa
d\tilde{s}^2=-\Bigl({u-(1-p)v\over u-(1+p)v}\Bigr)^{\pm \alpha}dudv
&+&{1\over 4}(u-(1-p)v)^{1\pm \alpha}(u-(1+p)v)^{1\mp \alpha}d\Omega^2,\\
\Phi&=&\Bigl({u-(1-p)v\over u-(1+p)v}\Bigr)^{\mp \alpha},
\eeqa
where $\alpha=1/\sqrt{2\omega +3}$.
Since the Ricci scalar takes the form
\beq
\tilde{R}=\Phi{R+3\nabla^2\Phi-{3\over 2}(\nabla\Phi)^2},
\eeq
and $R=8[(u-(1-p)v))(u-(1+p)v)]^{-2}p^2uv$, we
find that a singularity at $u=(1-p)v$ for $v>0$ and $u=(1+p)v$ for
$v<0$. In the following we shall assume
$1\pm \alpha \geq 0$ in order for the singularity to locate
at the center. As in the case of the Einstein gravity,
structure of the singularity depends on $p$.  It is easily found that
for $p>1$ it is timelike in the region $v<0$ and spacelike $v >0$,
null for $p=1$, timelike for $p<1$.

By calculating the change of area with respect to derivative along
advanced null coordinate, it is found that an apparent horizon is
located at
\beq
u={1-p^2\over 1\pm \alpha p} v,
\eeq
and that for $p<1$ no apparent horizon exists. Hence the parameter
$p$ corresponds to the control parameter in this example of
the critical behavior and $p=1$ the point of the criticality.

Calculation of the black hole mass $M$ is also straightforward.
It is done by calculating the square root of
the proper area of an apparent horizon.  After a little algebra
it is found that $M$ is given by
\beq
M ={(1\pm \alpha)^{(1\mp \alpha)/2}(1\mp \alpha)^{(1\pm \alpha)/2}
pv(1+p)^{(1\mp \alpha)/2}\over 4(-1 \pm \alpha p)} (p-1)^{(1\pm \alpha)/2}.
\eeq
Thus near the criticality $M$ behaves as
$M \propto (p-1)^{(1\pm \alpha)/2}$. It  is the power law
of the critical behavior with the critical exponent
\beq
\beta ={1\over 2} \pm {1\over 2\sqrt{2\omega +3}},
\eeq
which clearly depends on $\omega$
and certainly reduces to the result of Einstein theory in the limit
 $\omega \rightarrow \infty$.
 This result strongly indicates that the critical phenomena depend
 on the parameter $\omega$. Indeed, it will be confirmed in the subsequent
 numerical analysis. 

\section{Collapse of Brans-Dicke
scalar field in double-null coordinates}

Next we study numerically the collapse of the Brans-Dicke scalar field.
We treat the problem with the action Eq.(\ref{class1})
in terms of null initial value formulation.
Since in the null initial value formulation the grid points are tied
to ingoing light rays, overall size of the grid becomes smaller
as system evolves and the resolution improves. Adaptive mesh refinement
algorithm used by Choptuik is not always necessary.

We take the following line element
\beq
ds^2=-a(u,v)^2dudv +r(u,v)^2d\Omega^2.
\eeq
Double-null coordinates are not unique. There remain two
degrees of gauge freedom which correspond to redefining $u$ and $v$.
We fix one of them by requiring $u=v$ on axis $r=0$.
The remaining gauge freedom will be fixed by the initial condition.

The equations of motion that derived from Eq.(\ref{class1}) are
\beqa
rr_{uv}+r_ur_v+{1\over 4}a^2&=&0,\label{classeq1}
\\
a^{-1}a_{uv}-a^{-2}a_ua_v-r^{-2}r_ur_v-{1\over 4}r^{-2}a^2
+{1\over 4}\phi_u\phi_v&=&0,\label{classeq2}
\\
r\phi_{uv}+r_u\phi_v +r_v\phi_u&=&0.
\label{classeq3}
\eeqa
And constraint equations are
\beqa
r_{uu}-2a^{-1}a_ur_u+{1\over 4}r\phi_u^2&=&0,\\
r_{vv}-2a^{-1}a_vr_v+{1\over 4}r\phi_v^2&=&0,
\label{classeq4}
\eeqa
where $r_u \equiv \p r/\p u$, for example.
These equations are solved numerically.
Since the usual iteration scheme for solving these equations
does not work because of their nonlinearity, we
adopt the first order scheme developed by
Hamade and Stewart\cite{hs}.

We introduce the following new variables
\beq
d={a_v\over a}, f=r_u, g=r_v, p=\phi_u, q=\phi_v.
\eeq
The system of equations (\ref{classeq1}-\ref{classeq4}) can then be converted to
a first order system:
\beqa
f_v+r^{-1}(fg+{1\over 4}a^2)&=&0,\label{f}\\
d_u-r^{-2}(fg+{1\over 4}a^2)+{1\over 4}pq&=&0,\label{d}\\
p_v+r^{-1}(fq+gp)&=&0,\label{p}\\
q_u+r^{-1}(fq+gp)&=&0,\label{q}\\
g_v-2dg+{1\over 4}rq^2&=&0,
\label{g}
\eeqa
and the equations defining variables
\beqa
a_v-ad&=&0,\label{a}\\
r_v-g&=&0,\label{r}\\
\phi_v-q&=&0\label{ph}.
\eeqa

We have to specify the boundary conditions on axis, $u=v$.
Of course, $r=0$ there and this implies $r_u+r_v=0$, i.e., $f=-g$.
Eq.(\ref{f}) will be regular on axis if and only if $a=2g$.
Eq.(\ref{p}) implies $p=q$ on axis. The boundary value for $a$ and
$\phi$ are obtained by requiring $a_r=\phi_r=0$ there.

We need to give initial data $d$ and $q$ on the initial surface
$u=0$. The remaining gauge freedom mentioned in the second paragraph
is the choice of $d$ on $u=0$. This is arbitrary, and
we choose $d=0$. As for the initial condition
of $q$ (or $\phi$) on $u=0$, we typically take
\beq
\phi(u=0,v)=1+\phi_0v^2\exp(-(v-v_0)^2/v_1^2).
\label{scalar}
\eeq
We then integrate outward from axis Eq.(\ref{a}) to obtain $a$,
Eq.(\ref{r}) to obtain $r$, Eq.(\ref{g}) to obtain $g$,
Eq.(\ref{f}) to obtain $f$ and Eq.(\ref{p}) to obtain $p$.
Thus we set initial conditions.

The integration in the $u$-direction is done using
an explicit difference algorithm. The integration in the
$v$-direction is done using an
implicit algorithm, which is necessary to ensure stability.
However, the integration can be made
explicit. Details of algorithm are given in\cite{hs}.\footnote{
However, slight modification is necessary to ensure that the system
has second-order accuracy; Eq.(3.5) in their paper should be
replaced with $z_n={1\over 2}\bigl(\hat{z}_n+z_w+{h\over 2}
(G(\hat{y}_n,\hat{z}_n)
+G(y_w,z_w))\bigr)$. We thank T.Harada for pointing this out to us.}

In Fig.1, we plot the scalar field at the center with
marginally subcritical initial parameter as a function of
logarithm of
the central proper time $\tau\equiv -\log(t^*-t)$,
where $t^*$ is the time when
an infinitesimal black hole forms.
We find that the scalar
field oscillates with period $\Delta \simeq 3.43$ in $\tau$-coordinate,
in agreement with Choptuik.

Fig.2 shows the black hole mass for marginally supercritical evolution
as a function of $\phi_0-\phi_0^*$
with $\phi_0^*$ being the critical parameter. The least square fit
shows the exponent $\beta \simeq 0.38$, in agreement with Choptuik.

These results confirm the logical consistency  and accuracy of
our numerical code.

\section{critical phenomena in the brans-dicke theory of gravitation}

Now we study the behavior of a scalar field  for various values of
$\omega$, $1000 \geq 2\omega +3 > 0$ by conformally transforming the
Choptuik's solution reproduced in the previous section.
This is done by the relation (\ref{conf}).
The results are shown in Figures and Table.
Fig.3 shows the scalar filed at the center as a function
of logarithm of the central proper time $\tau\equiv -\log(t^*-t)$
for several values of $\omega$. The scalar field oscillates
greatly for smaller $\omega$.
Fig.4 shows the black hole mass for marginally supercritical evolution
as a function of initial parameter for $2\omega+3=2$.
The least square fit gives the exponent $\beta \simeq 0.43$.
Table  summarizes the numerical results. As can be seen
the exponent $\beta$ depends strongly on $\omega$, hence it
is not universal quantity. This is the first numerical evidence
of non-universality\cite{mason} for the collapse of single matter content.
On the other hand the scaling parameter
$\Delta$ depends weakly on $\omega$.

\vskip0.5cm

{\bf Table \quad} Brans-Dicke parameter,
critical exponents, and scaling parameters.

\begin{center}
\begin{tabular}{|c|c|c|}\hline
\makebox[50mm]{2$\omega$+3}&
\makebox[50mm]{$\beta$}&
  \makebox[50mm]{$\Delta$}    \\ \hline
\hline
 0.50 &
 \( {\displaystyle 0.50  } \) &  \( {\displaystyle 3.41 }\)   \\ \hline
 1.0 &
 \( {\displaystyle 0.46 } \) &  \( {\displaystyle 3.43 }\)   \\ \hline
 2.0 &
 \( {\displaystyle 0.43 } \) &  \( {\displaystyle 3.45 }\)   \\ \hline
 5.0 &
 \( {\displaystyle 0.41 } \) &  \( {\displaystyle 3.46 }\)   \\ \hline
 10 &
 \( {\displaystyle 0.38} \) &  \( {\displaystyle 3.46 }\)   \\ \hline
 50 &
 \( {\displaystyle 0.38 } \) &  \( {\displaystyle 3.47 }\)   \\ \hline
 200 &
 \( {\displaystyle 0.38 } \) &  \( {\displaystyle 3.47 }\)   \\ \hline
 1000 &
 \( {\displaystyle 0.38 } \) &  \( {\displaystyle 3.47 }\)   \\ \hline
\end{tabular}
\end{center}
\ \\

It is to be noted that critical phenomena in the Brans-Dicke theory does
emerge in discrete self-similar manner. Thus stability analysis around a
{\it continuous} self-similar solution gives no information for
the stability of the actual dynamical solutions.

\section{Summary}

We have studied the collapse of the Brans-Dicke scalar field to
examine the dependence of the critical phenomena on the theory of
gravitation.
First, by conformally transforming the Roberts's solution into
the Brans-Dicke frame, we find for $\omega > -3/2$ that
a continuous self-similarity continues and that
the critical exponent does depend on $\omega$.
Second, by conformally transforming the Choptuik's solution into
the Brans-Dicke frame, we find for $\omega > -3/2$ that
at the critical solution shows discrete self-similarity.
However, the critical exponent depens strongly on $\omega$, while 
the scaling parameter depends weakly  on $\omega$. 
This weak dependence on the theory of gravitation may justify 
the approximation employed by Price and Pullin.\cite{pp} 
They find that the critical solution can be well approximated by 
a flat spacetime scalar field solution. 
Numerical dynamical solutions show discrete self-similarity.
These are another examples of discrete self-similar critical solutions.

 Given these numerical examples, the critical phenomena are not
{\it universal} in the original sense. 
 The apparent analogy to statistical system
may be just analogy not fundamental. 
Rather, the phenomena numerically observed can be  regarded 
as a manifestation of the features of the infinite dimensional 
dynamical system. From this point of view, there are many interesting
problems to be solved. For example, 
``Does the discrete self-similarity corresponds to the limit cycle?''
or, more generally,  
``How can we classify the gravitational collapse phenomena from
the point of the dynamical system?'',
``Are there any examples which give rise to bifurcation phenomena?'' 

We hope to report on the resolution of these problems in the future.

\acknowledgments

T.C. would like to thank Dieter Mason
for useful discussions and Misao Sasaki for useful comment on
the conformal transformation. He is also grateful to H.Sato for
continuous encouragement.

\newpage
\vskip 0.3in
\centerline{FIGURE CAPTIONS}
\vskip 0.05in

\newcounter{fignum}
\begin{list}{Fig.\arabic{fignum}.}{\usecounter{fignum}}

\item The scalar field on the center with a marginally subcritical
initial parameter
as a function of logarithm of
the central proper time $\tau=-\log(t^*-t)$, where $t^*$ is the time when
an infinitesimal black hole forms.
In this time scale the scalar
field oscillates with a periodicity $\Delta \simeq 3.43$.

\item The mass of a formed black hole for supercritical solutions
as a function of $p-p^*$
with $p^*$ being the critical initial parameter. The least square fit
gives the exponent $\beta \simeq 0.38$.

\item The scalar field on the center for
$2\omega +3=0.5,1,2$
with a marginally subcritical
initial parameter
as a function of logarithm of
the central proper time $\tau=-\log(t^*-t)$.

\item The mass of a formed black hole for supercritical solutions
for $2\omega +3=2$
as a function of $p-p^*$.
The least square fit
gives the exponent $\beta \simeq 0.43$.

\end{list}


\begin{thebibliography}{99}

\bibitem{chop}
M.W. Choptuik,
{\it Phys.\ Rev.\ Lett.\/} {\bf 70} 9 (1992).

\bibitem{ae}
A.M. Abrahams and C.R. Evans,
{\it Phys.\ Rev.\ Lett.\/} {\bf 70} 2980 (1992).

\bibitem{ec}
C.R. Evans and J.S. Coleman,
{\it Phys.\ Rev.\ Lett.\/} {\bf 72} 1782 (1993).

\bibitem{cs}
T. Chiba and M. Siino, 
{\it Disappearance of Critical Behavior
in Semiclassical General Relativity}, {\it KUNS-1384}, 
submitted to CQG.

\bibitem{bd}
C.B. Brans and R.H. Dicke,
{\it Phys.\ Rev.\/} {\bf 124} 925 (1961).

\bibitem{rob}
M.D.Roberts, {\it Gen.\ Rel.\ Grav.\/} {\bf 21} 907 (1989);
Y.Oshiro, K.Nakamura, and A.Tomimatsu, {\it Prog.\ Theor.\ Phys.\/}
{\bf 91} 1265 (1994);
P.R.Brady, {\it Class.\ Quantum Grav.\/} {\bf 11} 1255 (1994).

\bibitem{hs}
R.S. Hamade and J.M. Stewart, 
{\it Class.\ Quantum Grav.\/} {\bf 13} 497 (1996).

\bibitem{mason}
D. Mason, {\it Phys.\ Lett.\/} {\bf B366} 82 (1996).

\bibitem{pp}
R.H.Price and J.Pullin, gr-qc/9601009.


\end{thebibliography}
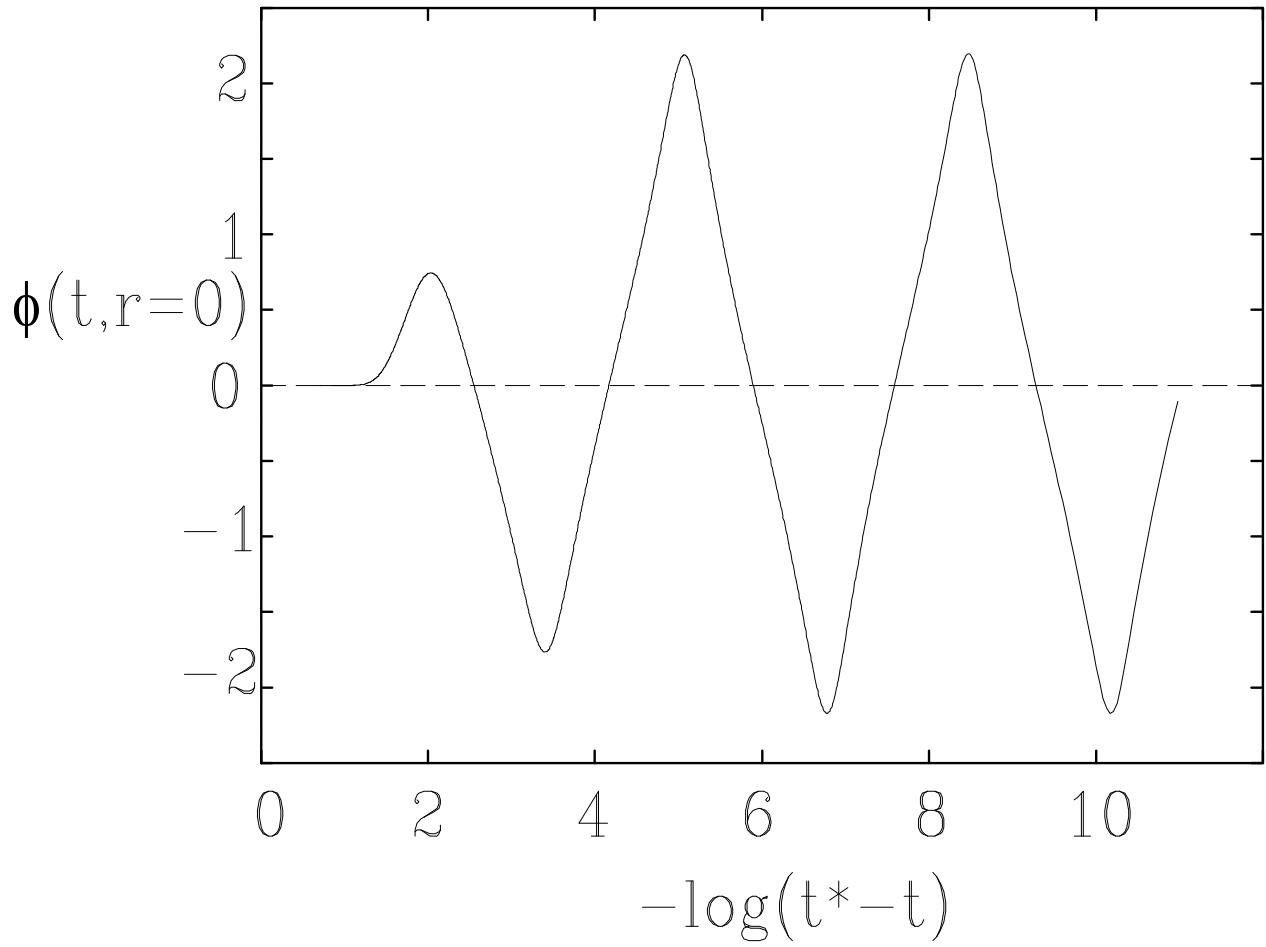
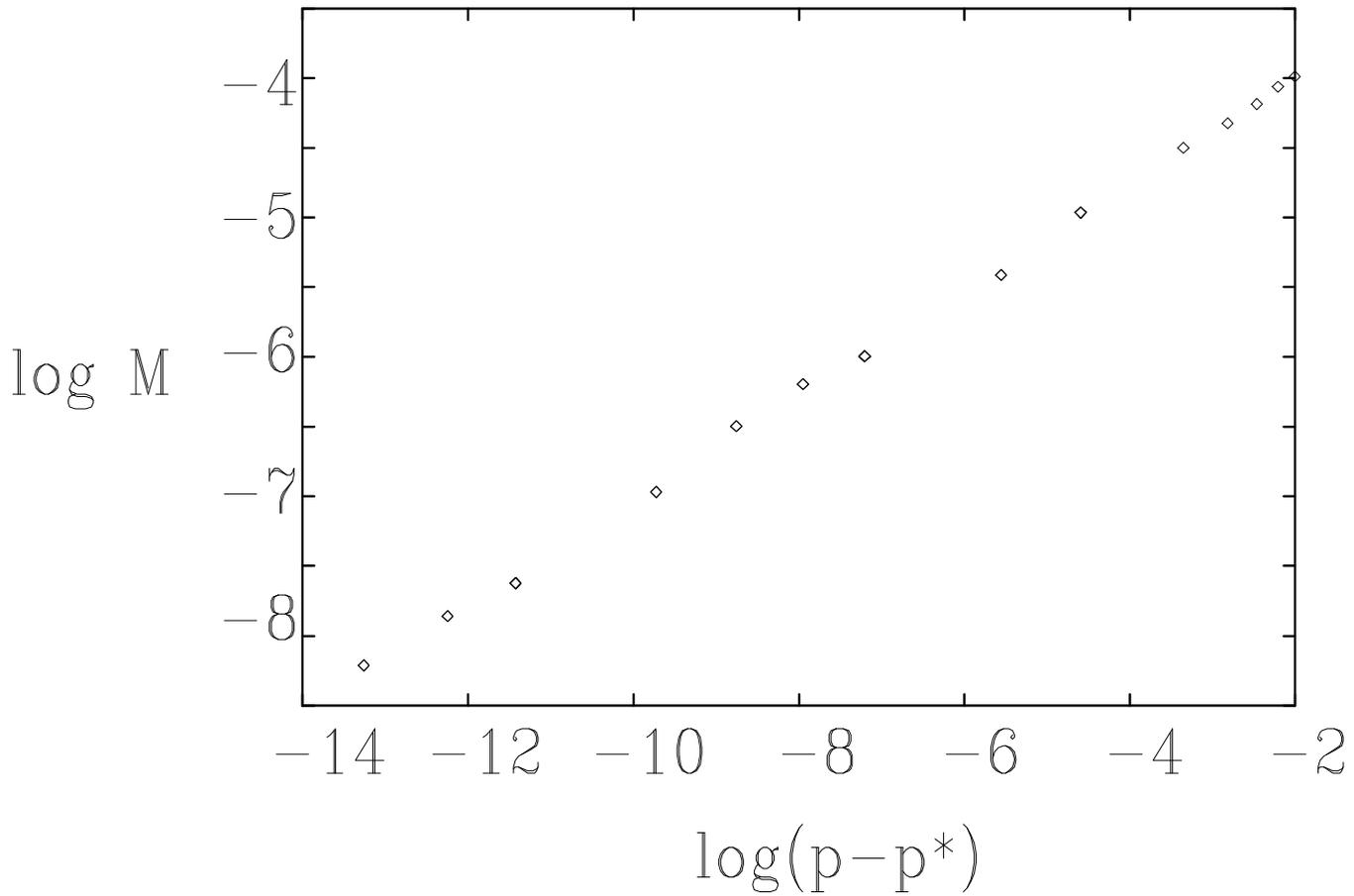
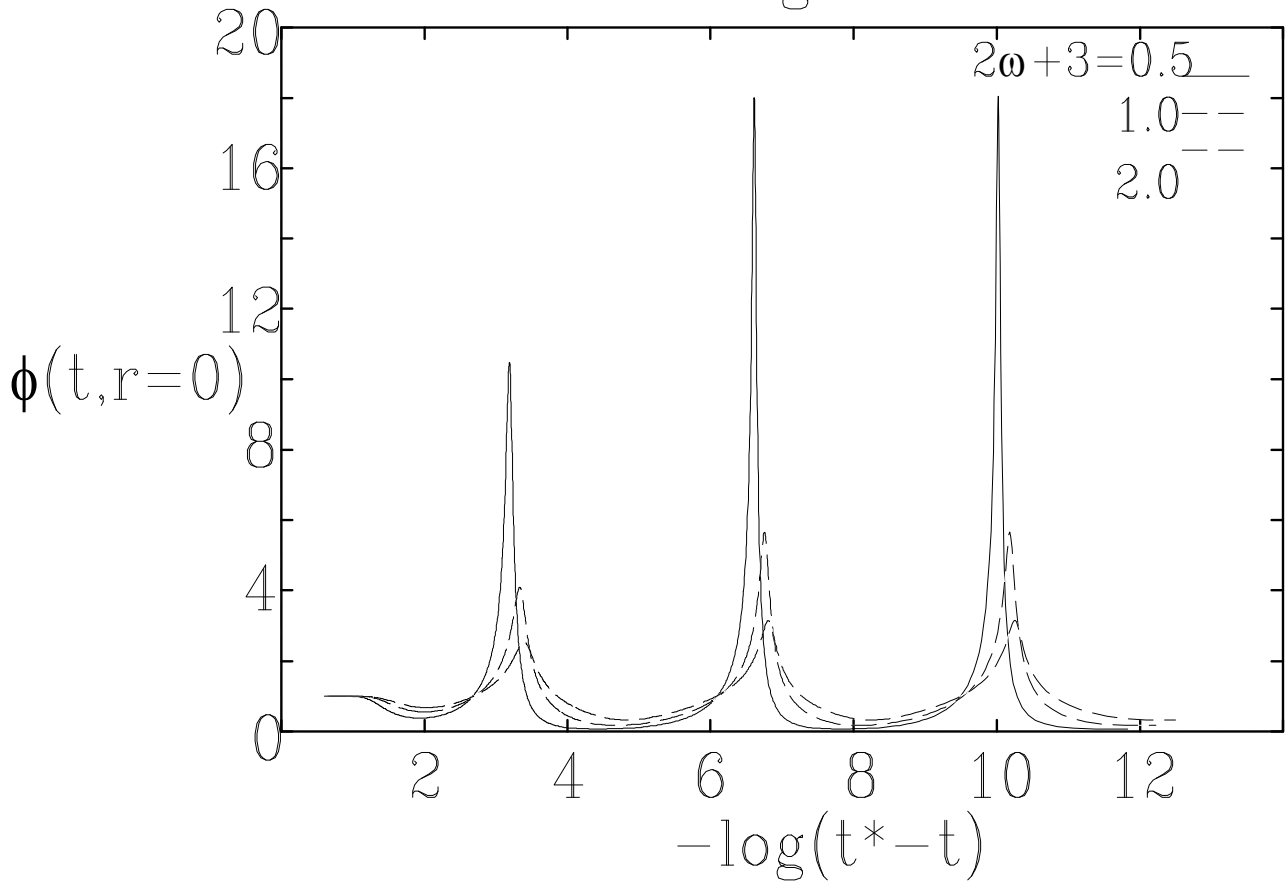
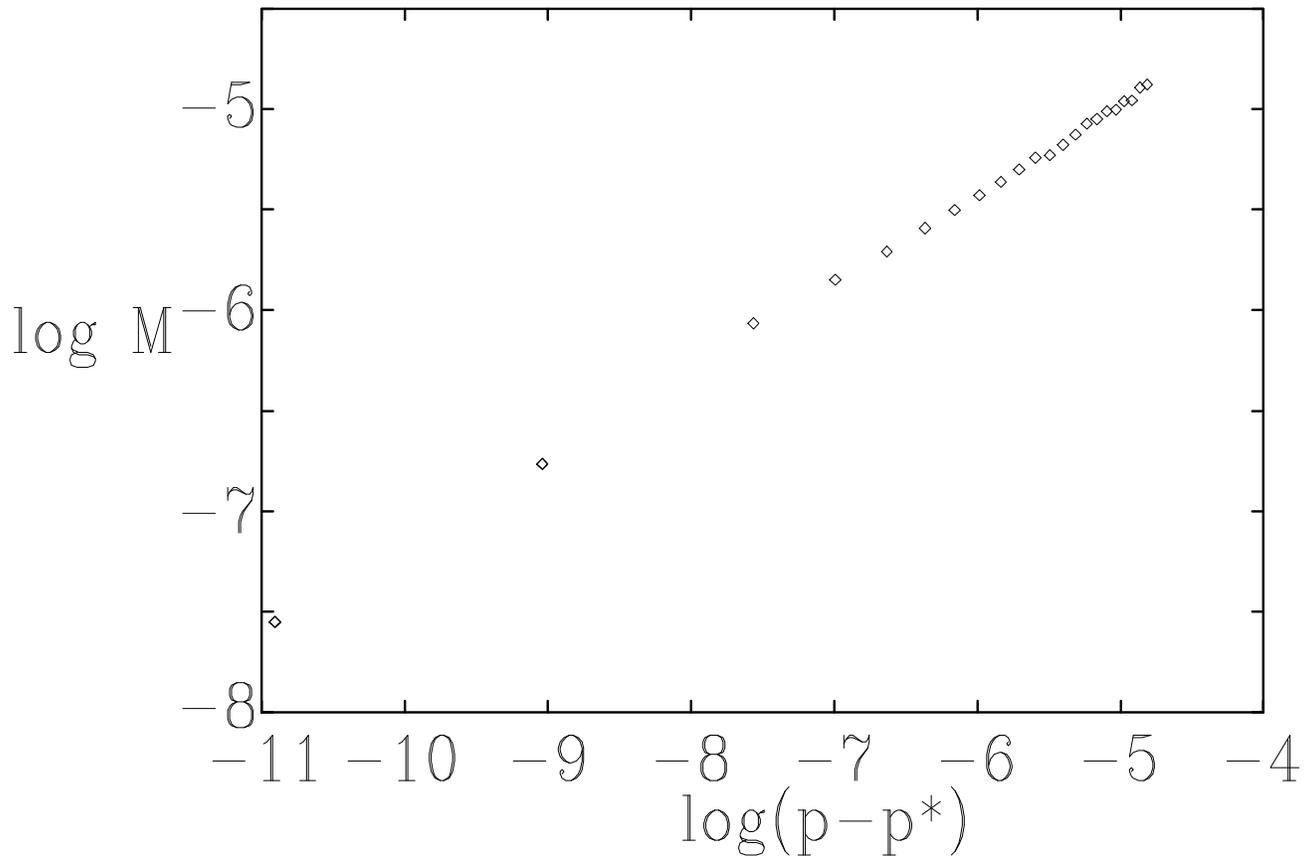
\end{document}